\documentclass[conference]{IEEEtran}

\IEEEoverridecommandlockouts

\usepackage{cite}
\usepackage{amsmath,amssymb,amsfonts}
\usepackage{graphicx}
\usepackage{textcomp}
\usepackage{xcolor}
\usepackage{algorithm}
\usepackage[noend]{algorithmic}
\usepackage{bm}
\usepackage{amsmath,amsfonts,amsthm}
\usepackage{mathtools}
\usepackage{makecell}
\usepackage{enumitem}
\usepackage{bm}
\usepackage{multirow}
\usepackage{url}
\usepackage{comment}
\usepackage{tikz}

\makeatletter
\newcommand{\linebreakand}{%
  \end{@IEEEauthorhalign}
  \hfill\mbox{}\par
  \mbox{}\hfill\begin{@IEEEauthorhalign}
}
\makeatother

\begin{document}

\DeclarePairedDelimiter\norm{\lVert}{\rVert}
\DeclarePairedDelimiter\abs{\lvert}{\rvert}
\newcommand{\Xin}[1]{{\textcolor{blue}{\bf[[Xin: #1]]}}}

\DeclareRobustCommand{\circled}[1]{%
  \tikz[baseline=(char.base)]{
    \node[shape=circle,draw,inner sep=1pt] (char) {\small #1};
  }%
}

\title{Mitigating Artifacts in Pre-quantization Based  Scientific Data Compressors with Quantization-aware Interpolation}

\author{
    \IEEEauthorblockN{Pu Jiao}
    \IEEEauthorblockA{\textit{University of Kentucky} \\
    Lexington, KY, USA \\
    pujiao@uky.edu}
    \and
    \IEEEauthorblockN{Sheng Di}
    \IEEEauthorblockA{\textit{Argonne National Laboratory} \\
    Lemont, IL, USA \\
    sdi1@anl.gov}
    \and
    \IEEEauthorblockN{Jiannan Tian}
    \IEEEauthorblockA{\textit{Oakland University} \\
    Rochester, MI, USA\\
    jtian@oakland.edu}
    \and
    \IEEEauthorblockN{Mingze Xia}
    \IEEEauthorblockA{\textit{Oregon State University} \\
    Corvallis, OR, USA \\
    xiami@oregonstate.edu}
    \linebreakand 
    \IEEEauthorblockN{Xuan Wu}
    \IEEEauthorblockA{\textit{Oregon State University} \\
    Corvallis, OR, USA \\
    wuxuan@oregonstate.edu}
    \and
    \IEEEauthorblockN{Yang Zhang}
    \IEEEauthorblockA{\textit{Miami University} \\
    Oxford, OH, USA \\
    zhang981@miamioh.edu}
    \and
    \IEEEauthorblockN{Xin Liang\IEEEauthorrefmark{1}\thanks{\IEEEauthorrefmark{1}Corresponding author: Xin Liang, School of Electrical Engineering and Computer Science, Oregon State University, OR 97331, USA.}}
    \IEEEauthorblockA{\textit{Oregon State University} \\
    Corvallis, OR, USA \\
    lianxin@oregonstate.edu}\\
    \and
    \IEEEauthorblockN{Franck Cappello}
    \IEEEauthorblockA{\textit{Argonne National Laboratory} \\
    Lemont, IL, USA \\
    cappello@mcs.anl.gov}
}


\maketitle

\begin{abstract}
Error-bounded lossy compression has been regarded as a promising way to address the ever-increasing amount of scientific data in today's high-performance computing systems. 
Pre-quantization, a critical technique to remove sequential dependency and enable high parallelism, is widely used to design and develop high-throughput error-controlled data compressors. 
Despite the extremely high throughput of pre-quantization based compressors, they generally suffer from low data quality with medium or large user-specified error bounds. 
In this paper, we investigate the artifacts generated by pre-quantization based compressors and propose a novel algorithm to mitigate them. Our contributions are fourfold: (1) We carefully characterize the artifacts in pre-quantization based compressors to understand the correlation between the quantization index and compression error; (2) We propose a novel quantization-aware interpolation algorithm to improve the decompressed data; (3) We parallelize our algorithm in both shared-memory and distributed-memory environments to obtain high performance; (4) We evaluate our algorithm and validate it with two leading pre-quantization based compressors using five real-world datasets. 
Experiments demonstrate that our artifact mitigation algorithm can effectively improve the quality of decompressed data produced by pre-quantization based compressors while maintaining their high compression throughput. 
\end{abstract}

\begin{IEEEkeywords}
Lossy compression, artifact mitigation, Euclidean distance transform, parallel computing.
\end{IEEEkeywords}

\section{Introduction}

Today's high-performance computing (HPC) systems are generating scientific data at unprecedented amount and speed. 
For example, a single high-resolution climate simulation can produce tens of terabytes of data every 16 seconds with pre-exascale systems~\cite{foster2017computing}, and this pressure is expected to intensify due to the recent deliveries of exascale systems~\cite{aurora, frontier, elcaptain}.
This growth will cause significant challenges for underlying data management tasks, as advancements in networks and file systems cannot keep pace with computational throughput, thereby hindering scientific discoveries. 



Error-bounded lossy compression has emerged as a promising way to address the big data challenge in the scientific data management community. 
It effectively reduces data volumes by orders of magnitude, enabling efficient data storage, transfer, and in-situ processing. 
Meanwhile, it provides guaranteed error control for the decompressed data based on user specification, ensuring the integrity and fidelity of scientific data for post hoc analytics. 

With the increasing heterogeneity in the design of HPC systems, GPU-based error-bounded lossy compression is attracting growing attention, where pre-quantization becomes a vital technique to break the sequential dependency in CPU-based algorithms to ensure high parallelism.
First proposed in cuSZ~\cite{tian2020cusz}, pre-quantization applies linear-scaling quantization as a preprocessing step, making it the only stage in the compression pipeline that introduces errors. 
This removes the heavy sequential dependencies in major decorrelation methods (e.g., Lorenzo prediction~\cite{ibarria2003out}), allowing concurrent execution of both decorrelation and the following steps. 
As such, pre-quantization has been widely adopted in many other GPU-based compressors, such as FZ-GPU~\cite{zhang2023fz} and cuSZp~\cite{huang2023cuszp, huang2024cuszp2}, to achieve high compression throughput, as well as CPU-based compressors to enable homomorphic compression~\cite{agarwal2024szops} and accelerate MPI collectives~\cite{huang2024hzccl}. 

Despite the wide usage of pre-quantization based lossy compressors, their decompressed data may suffer from severe visual artifacts. 
Artifacts represent noticeable distortions in the reconstructed
data that lead to obvious visual differences or wrong interpretations. 
They are typically present in decompressed data when relatively large error bounds are used to achieve high compression ratios and throughput. 
As such, scientists are reluctant to adopt large error bounds in lossy compression, thereby limiting compression efficiency.

While artifact mitigation has been studied for a long time in the image compression community, extending it to scientific data compression is non-trivial due to the following reasons. 
First, scientific data are usually multidimensional floating-point numbers, which are more difficult to process than the integer-valued pixels typical in natural images.
Second, scientific data compression requires error control on the reconstructed data to ensure integrity, but traditional artifact mitigation techniques such as filtering~\cite{kundu1995enhancement, chen2001adaptive} and dithering~\cite{bhagavathy2009multiscale} may easily introduce significant errors that break such constraints. 
Finally, the artifact mitigation procedure should be relatively fast to meet application requirements, making computationally expensive methods such as deep neural networks~\cite{yeh2021deep, maleki2018blockcnn} and generative models~\cite{gen_ai_artifact} impractical. 


In this paper, we propose a novel post-decompression enhancement algorithm that can be  applied to data reconstructed by any pre-quantization based lossy compressor to significantly improve data quality. Since the algorithm operates entirely on decompressed data, it introduces no additional cost to the original compression or decompression process. To achieve this, we first conduct a detailed characterization of compression artifacts commonly observed in pre-quantization based lossy compressors, and then design an efficient quantization-aware interpolation algorithm based on this analysis. Most importantly, our approach is broadly applicable to a wide range of compressors, including SZp~\cite{agarwal2024szops, huang2024hzccl}, cuSZ~\cite{tian2020cusz}, cuSZp and cuSZp2~\cite{huang2023cuszp, huang2024cuszp2}, and FZ-GPU~\cite{zhang2023fz}. In summary, our contributions are as follows.


\begin{itemize}[leftmargin=*]
    \item We characterize the spatial patterns of the compression artifacts produced by pre-quantization. Specifically, we find that the change of quantization indices usually corresponds to a change of signs in the compression errors, and the magnitude of the compression error in a specific data point is generally proportional to its distance to the nearest quantization boundary. 
    \item We propose a novel quantization-aware interpolation algorithm based on the above characterization to mitigate artifacts post-decompression. As our approach does not touch the compression stage, it maintains the high throughput of pre-quantization based compressors while significantly improving the quality of their decompressed data.
    \item We parallelize our algorithm in shared-memory and distributed-memory environments with OpenMP and MPI, respectively. We further optimize the distributed-memory parallelization to eliminate the artifacts near processor boundaries while maintaining high scalability.
    \item We validate our algorithm with two leading pre-quantization based compressors, cuSZ~\cite{tian2020cusz} and cuSZp2~\cite{huang2024cuszp2}, using five real-world scientific datasets. Experimental results demonstrate that our approach leads to up to \textbf{108.33\%} improvement in the Structural Similarity Index Measure (SSIM), a widely recognized measurement for perceived quality.   This corresponds to up to \textbf{1.17}$\bm{\times}$ and \textbf{1.34}$\bm{\times}$ gain in compression ratios for cuSZ and cuSZp, respectively, when SSIM is set to the same level. Our approach can also enforce a relaxed error bound with improved Peak Signal-to-Noise Ratio (PSNR), which is very beneficial for scientific use cases requiring high compression throughput and quality.
\end{itemize}

The rest of the paper is organized as follows. Section~\ref{sec:related} discusses the related work. Section~\ref{sec:background} provides the necessary background on pre-quantization and Euclidean distance transform, which will be referenced in the later sections. 
Section~\ref{sec:overview} presents an overview of our algorithm. 
The artifact mitigation methods are introduced in Section~\ref{section:characterization}, and the detailed implementation and parallelization are described in Section~\ref{section:method} and Section~\ref{section:parallelization}. 
Section~\ref{section:evaluation} presents and analyzes the experimental evaluations, and Section~\ref{section:conclusion} concludes this paper with a vision for future work.

\section{Related Work}
\label{sec:related}
In this section, we review related work on error-controlled lossy compressors for scientific data and artifact mitigation techniques for image compression. 

\subsection{Lossy Compression for Scientific Data}

Due to the ever-increasing size of scientific data, error-bounded lossy compression has become a cornerstone of scalable scientific computing. 
It addresses the limited compression ratios of generic lossless compressors~\cite{gzip, zstd, blosc} and unbounded errors of traditional lossy compressors~\cite{wallace1992jpeg, rabbani2002jpeg2000} simultaneously, and has been widely used to reduce the volume of simulation outputs for storage and I/O~\cite{liang2019improving}, accelerate data transfer~\cite{liu2023optimizing}, and support real-time processing in streaming applications~\cite{underwood2023roibin}. Moreover, some recent studies further extend lossy compression to preserve topological features, enabling better support for topology-sensitive scientific analysis tasks~\cite{msz_yuxiao,li2025multitier,xia2024preserving,xia2025tspsz,xia2025time,nathanieltfz}.

Error-bounded lossy compressors are typically categorized into two families: prediction-based and transform-based, depending on the methods they use to decorrelate data. 
ISABELA~\cite{lakshminarasimhan2013isabela} is one of the first prediction-based scientific data compressors, and it leverages sorting and B-spline fitting to perform data decorrelation. 
Later, SZ~\cite{liang2022sz3} extends this idea by incorporating more advanced predictors such as Lorenzo~\cite{ibarria2003out}, regression~\cite{sz18}, and splines~\cite{zhao2021optimizing}, and combines them with linear-scaling quantization~\cite{sz17} to enforce guaranteed error control. 
ZFP~\cite{lindstrom2014fixed} is a representative transform-based compressor that relies on block-wise domain transforms and embedded encoding. It is one of the fastest scientific data compressors due to its concise design and careful optimizations. 
TTHRESH~\cite{ballester2019tthresh} is another transform-based compressor based on singular value decomposition, but its throughput is relatively low due to the cost of singular value decomposition.
MGARD~\cite{ainsworth2018multilevel,ainsworth2019multilevel} and SPERR~\cite{li2023lossy} are two other popular scientific data compressors featuring adaptive resolution and progressive representations. 

With the recently delivered GPU-based exascale computing systems~\cite{aurora, frontier, elcaptain}, there is a pressing need to develop high-performance scientific data compressors with advanced GPUs to keep up with the data generation speed. 
However, sequential dependencies in mainstream prediction-based compressors~\cite{lakshminarasimhan2013isabela, lindstrom2014fixed, liang2022sz3} severely limit their performance in GPUs. 
Pre-quantization was proposed first in cuSZ~\cite{tian2020cusz} to break this barrier, which performs linear-scaling quantization as a preprocessing step to ensure high parallelism in Lorenzo prediction. 
FZ-GPU~\cite{zhang2023fz} extends this idea by designing and incorporating new encoding algorithms, which further improves the compression throughput. 
Recently, cuSZp2~\cite{huang2023cuszp, huang2024cuszp2} has been proposed to combine pre-quantization with a simple prediction method and optimized memory access, achieving over 300 GB/s end-to-end compression throughput on NVIDIA A100 GPUs. 
Furthermore, SZp~\cite{agarwal2024szops} uses pre-quantization to support fast operations on compressed data, and HZ-CCL~\cite{huang2024hzccl} employs it to accelerate MPI\_Allreduce collectives.




\subsection{Artifact Mitigation}
Compression artifacts have been observed and studied in the image compression community for decades. 
Image compression artifacts are broadly categorized as blocking, ringing, blurring, and texture deviation, and they can be mitigated with different techniques. 
Image enhancement methods such as spatial filters~\cite{kundu1995enhancement, chen2001adaptive} apply lowpass filters on the decompressed images to improve the quality, which reduces blocking and ringing artifacts. 
Image restoration methods formulate artifact mitigation as an image recovery problem, and leverage prior knowledge such as statistical assumptions and compression constraints to restore the images using criterion-based methods~\cite{hong1996practical, wu1992improved} or constraint-based methods~\cite{lai1996image, youla1978generalized}.

With the recent advancement in deep learning and artificial intelligence, there is a growing trend in applying deep neural networks~\cite{yeh2021deep, maleki2018blockcnn} and vision transformers~\cite{fan2022sunet,tsai2022stripformer}. Although these methods have been proven effective for images, they cannot be directly applied to scientific data  because these methods prioritize visual quality and often overlook numerical error constraints.
This may result in unbounded errors in the decompressed data, leading to uncertainties in downstream scientific visualization and analytics. 
In addition, most of these algorithms target image compression artifacts produced by JPEG, but scientific data compressors employ various compression methods that are significantly different from those used in JPEG. 
Furthermore, some of these methods are very time-consuming to train or perform, making them less ideal for scientific workflows that require high performance and scalability.

While compression errors have been studied in~\cite{lindstrom2017error}, the investigation only focuses on statistical information, such as error distributions and autocorrelation. 
A recent work~\cite{hipc23_jiao} has categorized artifacts in scientific data compressors into three types, but provides no solutions to address them. 
The most relevant work~\cite{daoce_sc24} leverages a Bézier curve to mitigate blocking artifacts. However, those blocking artifacts are an artifact of the authors' specific design to divide the data into separate blocks and compress each block with distinct error bounds, which do not represent the artifacts in generic compressors.

In this work, we fill the gap by first characterizing the artifacts in pre-quantization based scientific data compressors and then mitigating them using a novel quantization-aware interpolation method. 
We further parallelize our approach in both shared-memory and distributed-memory environments to achieve high throughput and scalability. 
Our algorithm is general and can be applied as a postprocessing stage during decompression for any pre-quantization based compressors. 
This preserves the high throughput of these compressors during compression while significantly improving the data quality for post hoc data analytics.

\section{Background}\label{sec:background}
In this section, we introduce the background of the pre-quantization technique and Euclidean distance transform algorithm. 
The former is the key step in pre-quantization based scientific data compressors, and the latter is the fundamental method used in the proposed artifact mitigation framework. 

\subsection{Pre-quantization in Scientific Data Compression}
Pre-quantization is a common technique to remove sequential dependency and improve parallelism in prediction-based scientific data compressors. 
The key idea is to quantize the input scientific data with the user-specified error bound in the beginning, and then change the subsequent prediction and encoding stages to a lossless manner. 
This directly generates the decompressed values for all data points, eliminating the need to reconstruct prior values before predicting subsequent points~\cite{sz17, sz18}.

We formally introduce pre-quantization as follows. 
Given a user-specified absolute error bound~$\epsilon$ and input data $\mathbf{D} = \{d_1, \dots, d_n\}$, pre-quantization maps each floating-point value $d_i$ to an integer index $q_i$ using:
\begin{equation}\label{eq:quant}
\small
q_i = \texttt{round}\left(\frac{d_i}{2\epsilon}\right),
\end{equation}
where \texttt{round}$(\cdot)$ indicates the rounding operator. 
During decompression, the data can be recovered as $d'_i = 2q_i\epsilon$, automatically providing the required error control as $\abs{d_i - d'_i} \leq \epsilon$. 
In other words, all data values falling in the interval $[(2q_i-1)\epsilon, (2q_i+1)\epsilon]$ will be quantized to the same integer $q_i$. 
As such, we define $[(2q_i-1)\epsilon, (2q_i+1)\epsilon]$ as the \textit{quantization interval} of $q_i$. 
Note that we also use quantized data or decompressed data to refer to $\{d'_1, \dots, d'_n\}$, and we define \textit{quantization error} as the difference between the original and quantized data, i.e., $d_i - d'_i$ for the $i$-th data point.

After pre-quantization, the resulting quantization index array $\mathbf{Q} = \{q_1, \dots, q_n\}$ is predicted and encoded in a lossless fashion. 
In cuSZ~\cite{tian2020cusz}, the multidimensional Lorenzo predictor~\cite{ibarria2003out} and Huffman encoder~\cite{huffman1952method} are used to achieve decent compression ratios; in cuSZp~\cite{huang2024cuszp2}, prediction is performed using only one prior data value and fixed-length encoding is employed to obtain high throughput.

\subsection{Euclidean Distance Transform}

The Euclidean Distance Transform (EDT) is a fundamental operation in image processing and computational geometry community~\cite{ultsch2022euclidean}. Given a binary image $I: \mathbb{Z}^k \to \{0,1\}$ over a $k$-dimensional grid of shape $n_1 \times \cdots \times n_k$, the EDT computes, for each background point (where $I(\mathbf{x}) = 0$), the Euclidean distance to the nearest foreground point (where $I(\mathbf{y}) = 1$). That is, for each $\mathbf{x} \in \mathbb{Z}^k$ such that $I(\mathbf{x}) = 0$, the EDT value is defined as:
\begin{equation}
\small
\text{EDT}(\mathbf{x}) = \min_{\mathbf{y} \in \mathbb{Z}^k,\ I(\mathbf{y})=1} \| \mathbf{x} - \mathbf{y} \|_2.
\end{equation}

Efficient algorithms exist for computing the exact EDT in linear time with respect to the number of elements in the domain~\cite{goutsias_general_2002, maurer_linear_2003, fabbri_2d_2008}. These methods operate in a dimension-by-dimension fashion. 
For example, Algorithm~\ref{alg:1dedt}, proposed by Maurer et al.~\cite{maurer_linear_2003}, iterates each point in a given row of data with 2 for-loops to construct and query partial Voronoi diagrams. The Voronoi sites are the boundary points in the binary image.  Initially, $l$ stores the count of candidate foreground points in the current row. In the first for-loop (line 2 to line 14), $g$ collects the boundary points and $h$ collects the corresponding indices of those Voronoi sites in the current dimension. The \textit{while} loop calls \textsc{RemoveEDT} to compare the current boundary point with previously collected boundary points to determine whether to discard any of them. \textsc{RemoveEDT}($g_{l-1}, g_l, f_i, h_{l-1}, h_l$)  returns true when $( c g_l - b g_{l-1} - a f_i - abc )>0 $, where $a = h_l - h_{l-1}$, $b= i - h_l$ and $c = i - h_{l-1}$. The second for-loop (line 19 - line 24) updates distances for all the points in the row using the kept boundary points. When Algorithm~\ref{alg:1dedt} is applied to the 2D case, it performs a first pass along each column independently to compute intermediate vertical distances to the nearest foreground pixels. Then, in a second pass, each row is processed using the column-wise results to yield the full 2D Euclidean distances. The algorithm generalizes to higher dimensions and allows straightforward parallelization across independent slices or dimensions; its computational complexity is linear in the number of data points.

\begin{algorithm}
\caption{VoronoiEDT}
\label{alg:1dedt}
\scriptsize
\renewcommand{\algorithmiccomment}[1]{\hfill\textcolor{gray}{// #1}}
\begin{flushleft}
\textbf{Input}: Initialized distance array $D$ with non-boundary location set to $\infty$, the size $n_d$ of $n$th dimension.
\\\textbf{Output}: Updated distance array $D$. 
\end{flushleft}
\begin{algorithmic}[1] 
\STATE $l \gets 0 \quad$  \COMMENT{Construct partial Voronoi diagrams}
\FOR{$i = 1$ to $n_d$}
    \STATE $x_i \gets (j_1, \dots, j_{d-1}, i, j_{d+1}, \dots, j_k)$
    \IF{$(f_i \gets D(x_i))\neq \infty$}
        \IF{$l < 2$}
            \STATE $l \gets l + 1$,  $g_l \gets f_i$, $h_l \gets i$
        \ELSE
            \WHILE{$l \geq 2$ and \textsc{RemoveEDT} ($g_{l-1}, g_l, f_i, h_{l-1}, h_l$)}
                \STATE $l \gets l - 1$
            \ENDWHILE
            \STATE $l \gets l + 1$, $g_l \gets f_i$, $h_l\gets i$
        \ENDIF
    \ENDIF
\ENDFOR
\IF{$(n_s \gets l) = 0$}
    \RETURN
\ENDIF
\STATE $l \gets 1 $ \quad \COMMENT{Query the Voronoi diagram }
\FOR{$i = 1$ to $n_d$}
    \WHILE{$l < n_s$ and $g_l + (h_l - i)^2 > g_{l+1} + (h_{l+1} - i)^2$}
        \STATE $l \gets l + 1$
    \ENDWHILE
    \STATE $D(x_i) \gets g_l+ (h_l - i)^2$
\ENDFOR
\RETURN
\end{algorithmic}
\end{algorithm}



\section{Overview}\label{sec:overview}
We formulate our research problem in this section, followed by an overview of the proposed framework. 
\subsection{Problem Formulation}

With a user-specified error bound $\epsilon$, error-bounded  scientific data compressors compress a scientific dataset $\mathbf{D} = \{d_1,\dots, d_n\}$ with $n$ data points into a binary format $\mathbf{C}$ with reduced size, which can be decompressed to a new representation $\mathbf{D'}= \{d'_1,\dots, d'_n\}$ with guaranteed $L^{\infty}$ error control, i.e., $\norm{\mathbf{D} - \mathbf{D'}}_\infty \leq \epsilon$ (equivalently, $\abs{d_i - d'_i} \leq \epsilon, \forall i \in \{1, \dots, n\}$). 
The goal of scientific data compressors is to reduce the size of $\mathbf{C}$ as much as possible while maintaining reasonable throughput for both compression and decompression under this given error bound. 

Since artifact mitigation operates after decompression, it improves reconstruction quality  without reducing compression efficiency, though it incurs certain post-decompression overhead.
Nonetheless, this is advantageous in many scientific use cases, such as data streaming~\cite{underwood2023roibin}, where compression throughput is required to keep up with data generation speed, and additional overhead after decompression is acceptable for post hoc data analytics.   
We define artifact mitigation in the context of scientific data compression as finding a transform $\mathcal{A}: \mathbb{R}^n \to \mathbb{R}^n$, which is applied to the decompressed data $\mathbf{D}'$ to generate a new representation $\mathcal{A}(\mathbf{D}')$ that is more similar to the original data $\mathbf{D}$ subject to \textit{relaxed error control}. This can be formulated as:
\begin{align*}
\max_{\mathcal{A}} \texttt{similarity}(\mathbf{D}, 
\mathcal{A}(\mathbf{D}')) \\
subject\ to \ \norm{\mathbf{D} - \mathcal{A}(\mathbf{D}')}_{\infty} \leq \epsilon',    
\end{align*}
where $\epsilon'$ is a relaxed error bound. Note that the decompressed data $\mathbf{D'}$ can be represented by $\{2q_1\epsilon, \dots, 2q_n\epsilon\}$ in the context of pre-quantization based scientific data compressors, where $q_i$ is the quantization index for the $i$-th input data $d_i$.

In this work, we primarily use the Structural Similarity Index Measure (SSIM)~\cite{wang2004image} to assess the similarity between the original and decompressed data, which is widely adopted in both the image compression and scientific data community. In particular, SSIM is defined as follows:
\begin{equation}
\small
SSIM(\mathbf{D_1}, \mathbf{D_2}) =  \frac{(2\mu_\mathbf{D_1}\mu_{\mathbf{D_2}} +c_1) (2\sigma_{\mathbf{D_1}\mathbf{D_2}}+c_2)}{(\mu_\mathbf{D_1}^2+\mu_{\mathbf{D_2}}^2+c_1) (\sigma_\mathbf{D_1}^2 +\sigma_{\mathbf{D_2}}^2 +c_2)},
\label{eqn:SSIM}
\end{equation}
where $\mu_\mathbf{D_1}$ and $\mu_{\mathbf{D_2}}$ are the sample means of the original data and decompressed data,  $\sigma_{D_1} ^2$ and $\sigma_{D_2}^2$ are the sample variances of the original data and the decompressed data, and $\sigma_{\mathbf{D_1}\mathbf{D_2}}$ is the sample covariance between the original data and the decompressed data.    Constants  $c_1 = 10^{-4}$, $c_2 = 9\times 10^{-4}$  used in this paper, following the calculation in Quick Compression Analysis Toolkit (QCAT) \cite{noauthor_szcompressorqcat_2024}.
Based on conventions, a sliding window is used to calculate SSIM in each region, and the average SSIM value is used to represent the overall SSIM for the dataset.
We use a default window size of 7 with a stride of 2 in our evaluation.

We use Peak Signal-to-Noise Ratio (PSNR) as a secondary measurement for the similarity, which is computed by: 
\begin{equation}
\small
PSNR(\mathbf{D_1}, \mathbf{D_2}) = 20  \log_{10}  \frac{ \max(\mathbf{D_1}) - \min(\mathbf{D_1})} { \sqrt{MSE(\mathbf{D_1}, \mathbf{D_2})} },
\label{eqn:PSNR}
\end{equation}
where $MSE(\mathbf{D_1}, \mathbf{D_2})$ is the mean squared error between $\mathbf{D_1}$ and $\mathbf{D_2}$. 
By combining PSNR and SSIM, the evaluation provides a comprehensive data quality assessment, capturing both numerical accuracy and perceptual similarity.

\subsection{Design Overview}
\begin{figure}
    \centering
\includegraphics[width=1\linewidth]{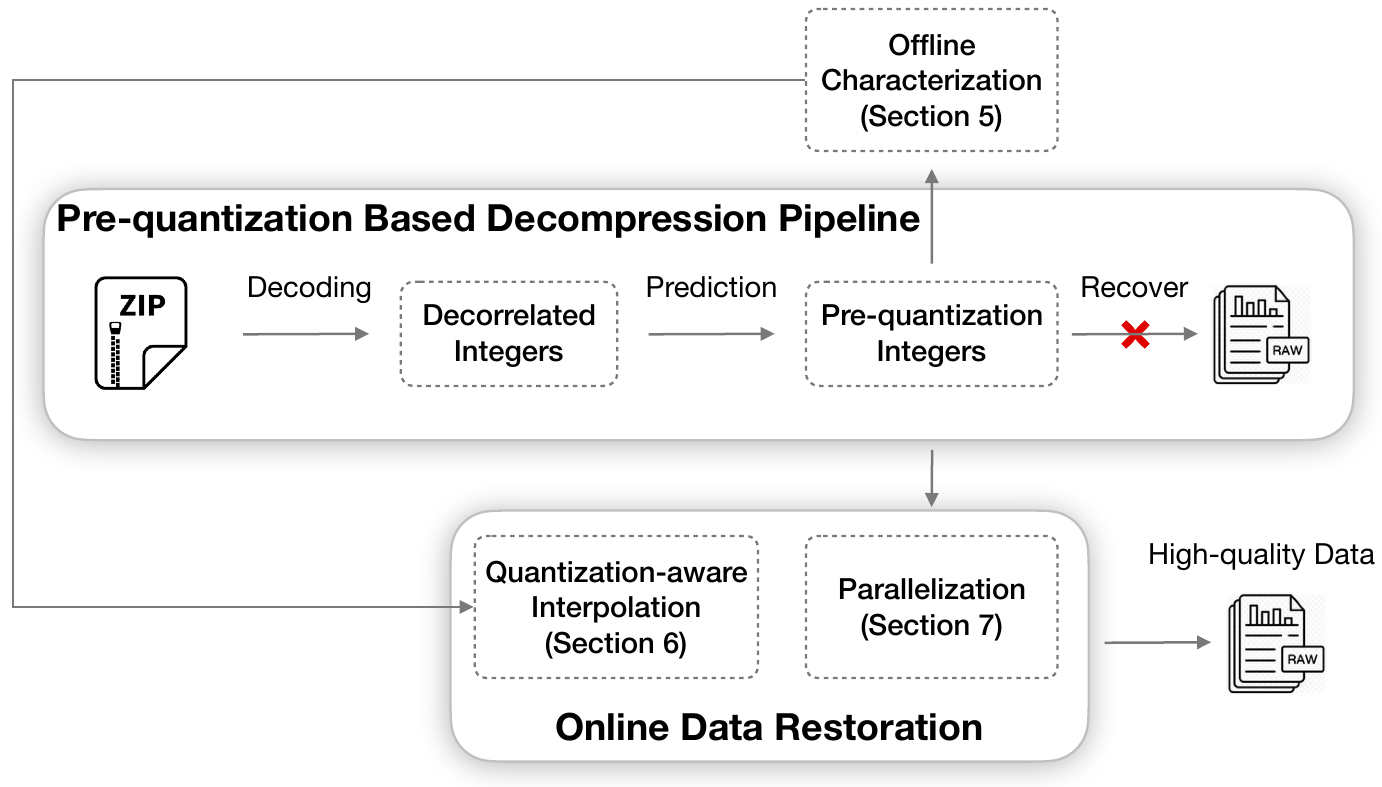}
    \caption{Overview of the proposed framework.}
    \label{fig:system-overview}
\end{figure}

We present an overview of the proposed framework in Fig.~\ref{fig:system-overview}, with the middle box depicting the generic decompression pipeline for pre-quantization based scientific data compressors. 
Since our framework does not touch the compression stages, we only discuss the workflow for decompression. 
In particular, we first conduct an offline characterization of the artifacts produced by pre-quantization based scientific data compressors to understand their correlations with the pre-quantization indices (Section~\ref{section:characterization}). 
Based on such prior knowledge, we develop an online data restoration framework, which accurately interpolates the compression errors using the pre-quantization indices (Section~\ref{section:method}) and adds them back to the decompressed data to obtain a high-quality data representation. 
We further optimize the online data restoration framework with OpenMP and MPI (Section~\ref{section:parallelization}), in order to achieve decent performance under different parallel scenarios.

\section{Characterization of Pre-quantization Artifacts}\label{section:characterization}

We characterize compression artifacts produced by pre-quantization in this section. In particular, we find that the signs and magnitudes of compression errors are strongly related to the distribution of quantization indices and we can leverage these relationships to guide error restoration during decompression.



\begin{figure}[h]
    \centering       
\includegraphics[width=0.9\linewidth]{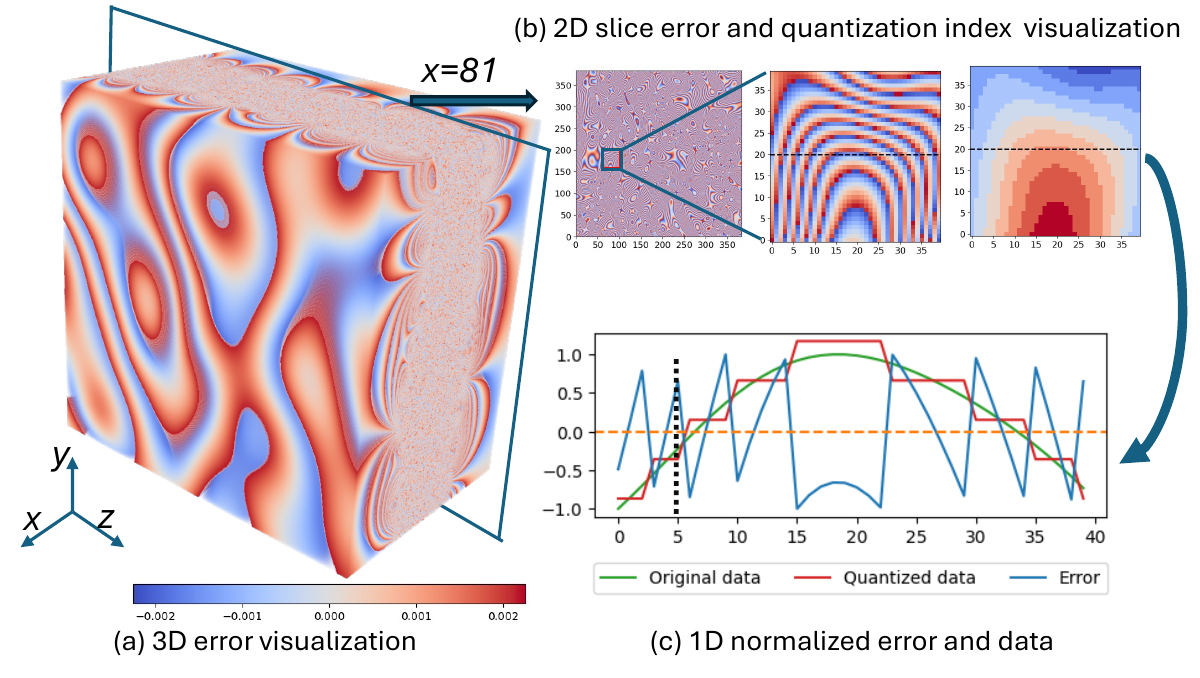}
        \caption{Visualization of errors in quantized data.}
    \label{fig:quantization_demo}
\end{figure}

We visualize the errors of pre-quantization in Fig.~\ref{fig:quantization_demo}, using the density field of the Miranda dataset~\cite{miranda} as an example. 
The relative error bound is set to $5\times 10^{-4}$, and similar patterns appear for other relative error bounds in our experiments. 
The left panel shows a 3D rendering of the volume, the top-right panels show zoomed 2D slices, and the bottom-right plot shows a 1D line cut from the 2D slice (normalized to $[-1, 1]$), with the original and quantized signals normalized to the original value range.

From the visualizations in Fig.~\ref{fig:quantization_demo},  we observe the following phenomena.

\textit{\underline{Clustering of quantization indices}}: We observe a clear clustering effect of quantization indices, which forms the contours in Fig.~\ref{fig:quantization_demo}(b). 
This is expected due to the local smoothness in scientific data: pre-quantization always converts values within an error bound to the same quantization index. 
We then define \textbf{quantization boundaries} as data points whose quantization index differs from at least one of its neighbors. 
By definition, quantization boundaries separate the data into different regions, where each region shares the same quantization index. 

\textit{\underline{Error sign flipping}}: According to Fig.~\ref{fig:quantization_demo}(b)(c), the signs of quantization errors flip at the quantization boundaries. 
This is because the data value is crossing from a quantization interval to another at the quantization boundaries. 
Assuming the transition is from $q$ to $q+1$ where the original data value is increasing from $2q\epsilon$ to $2(q+1)\epsilon$ (e.g., the black vertical line in Fig.~\ref{fig:quantization_demo}(c)), the quantization error would suddenly drop from near $+\epsilon$ to close to $-\epsilon$ when the value exceeds $(2q+1)\epsilon$, leading to the sign flipping. 
Furthermore, we notice that the error sign is correlated with the corresponding quantization index values: quantization indices at the lower boundaries have a positive sign, while those at the higher boundaries possess a negative sign. 
In this case, this sign corresponds to the sign of the forward-difference gradient of the quantization index.  

\textit{\underline{Error magnitude}}: 
Error magnitude tends to peak at quantization boundaries because values there are farthest from the reconstructed level. Between boundaries the errors vary smoothly with position owing to the underlying data smoothness, therefore interpolation can approximate the error field in non-boundary regions.

These observations lead to three core findings that underpin our compensation strategy in Section~\ref{section:method}, as detailed below.  

\noindent\textbf{(1) Sign at quantization boundaries}: The sign of the quantization error is correlated with the local gradient of the quantization index. When the gradient is positive, the error tends to be positive; when the gradient is negative, the error is generally negative. 

\noindent\textbf{(2) Error magnitude at quantization boundaries}: The magnitude of the quantization error is close to the error bound $\epsilon$ at quantization boundaries. 

\noindent\textbf{(3) Error magnitude at non-quantization-boundary regions}: The errors are continuous at non-quantization-boundary regions, which can be approximated by interpolation functions. 

    


\section{Quantization-aware Interpolation}\label{section:method}

At a high level, our method works as follows. Lossy compression introduces structured quantization errors that create visible posterization (banding) artifacts in the decompressed data. Because the quantization process is deterministic, the spatial pattern of these errors is predictable: errors are largest near quantization boundaries (where the quantization index changes) and diminish toward the midpoints between neighboring boundaries, where the error sign flips. Our algorithm exploits this structure by (1)~locating quantization boundaries and estimating the error sign and magnitude there, (2)~identifying where the error crosses zero (sign-flipping boundaries), and (3)~interpolating the error for every other point based on its distances to these two sets of boundaries. The interpolated error is then added back to the decompressed data, effectively smoothing out the banding artifacts while keeping the corrected values within a controlled error range.

\begin{figure}
    \centering
    \includegraphics[width=\linewidth]{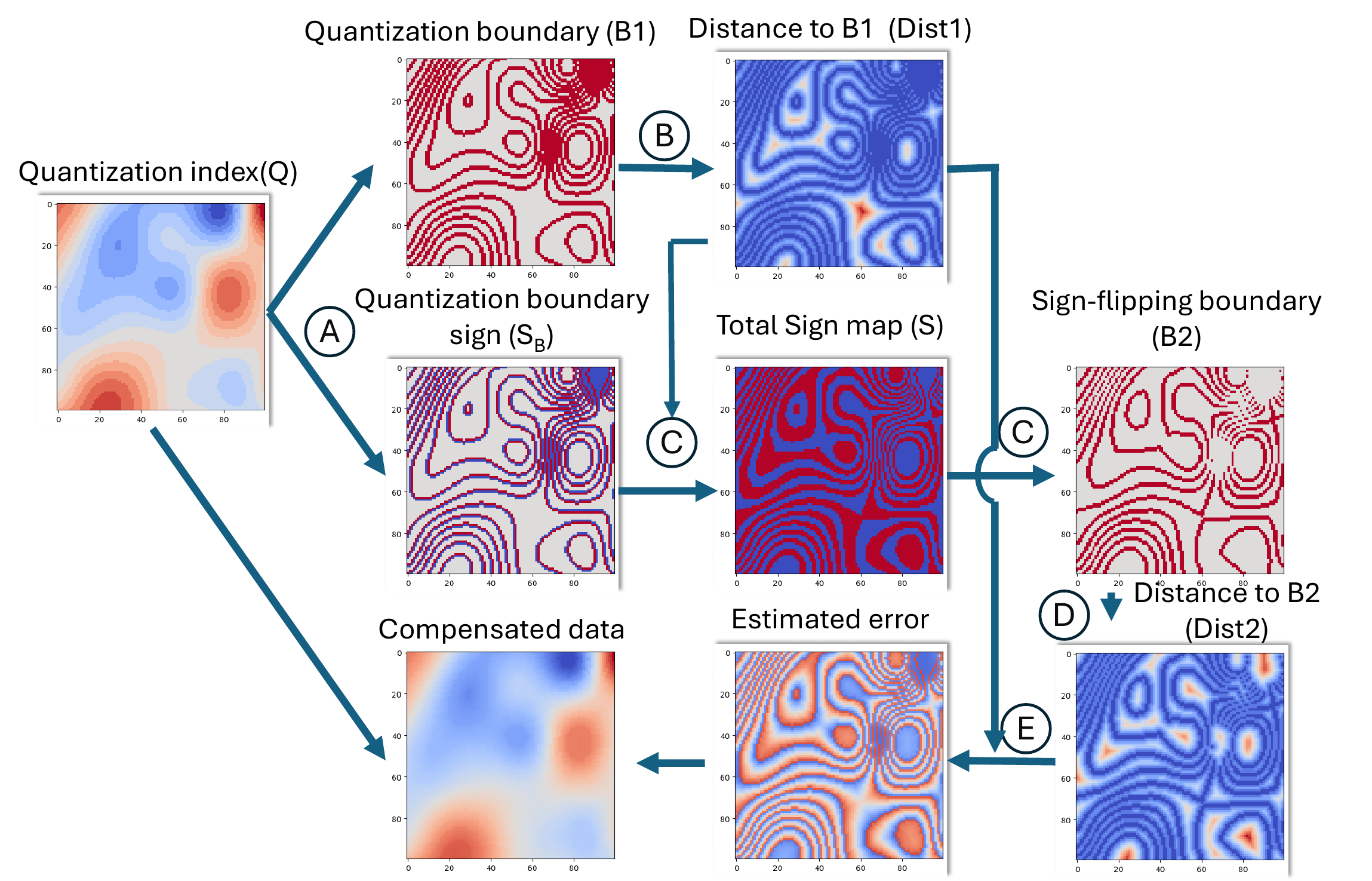}
    \caption{Workflow of the proposed quantization-aware interpolation and compensation using a 2D example. Step \circled{A} identifies quantization boundaries; step \circled{B} computes the first round EDT; step \circled{C} derives sign map and sign-flipping boundary; step \circled{D} computes the second round EDT; step \circled{E} performs interpolation and applies compensation.}
    \label{fig:workflow}
    \vspace{-0.5\baselineskip}
\end{figure}

The key steps of the algorithm are presented in Fig.~\ref{fig:workflow}.
In particular, we first identify the quantization boundaries and determine their error signs and magnitudes, which serves as the initial conditions for our interpolation (step \circled{A}). 
We then perform an EDT to compute each data point's distance to the quantization boundaries (step \circled{B}). 
After that, we propagate the sign to each data point from its closest quantization boundary, and then leverage the propagated signs to identify another set of sign-flipping boundaries, which are assumed to have an approximate value of $0$ (step \circled{C}).
We then perform another round of EDT to compute each data point's distance to the sign-flipping boundaries (step \circled{D}), and then perform interpolation for each point using their closest quantization boundary and sign-flipping boundary (step \circled{E}). 
In the following, we detail the procedure in each step.

\circled{A} {Identifying quantization boundaries} 
We present our algorithm for identifying quantization boundaries in Algorithm~\ref{alg:get_boundary_sign}, based on its definition and our findings in Section~\ref{section:characterization}.
In particular, we iterate every data point except those located on the domain boundary to check if it is a boundary (lines 3-17). 
If the current quantization index is equal to all its neighbors, we denote it as non-boundary and proceed to the next data point; otherwise, we mark it as a boundary and compute its sign (lines 7-11).
While the sign is mainly determined by the forward difference (line 9), we overwrite it with $0$ if the local gradient of any axis exceeds $1$ (lines 10-12). 
This prevents incorrect error estimation in fast-varying regions that violate our smoothness assumptions.
This step will produce the quantization boundary $\mathbf{B}_1$ and a sign map after excluding fast-varying regions, as shown in Fig.~\ref{fig:workflow}. 


\circled{B} {Computing distance to quantization boundaries (first round EDT)}
After we obtain the quantization boundary $\mathbf{B}_1$ from \circled{A}, we perform EDT (see Algorithm~\ref{alg:1dedt}) on $\mathbf{B}_1$ to obtain all data points' distance to $\mathbf{B}_1$ (denoted as $\mathbf{Dist}_1$) and their corresponding closest boundary data points (denoted as $\mathbf{I}_1$). 

\circled{C} {Estimating signs of errors and sign-flipping boundary} 
We will then propagate the signs from quantization boundaries to non-boundary regions using $\mathbf{I}_1$. 
As detailed in Algorithm~\ref{alg:propsigns}, the propagation can be done simply by iterating every data point and assigning it the sign of its closest boundary to get the total sign map $\mathbf{S}$.
After the signs are propagated to the whole domain, we can use \textsc{GetBoundary} to construct the sign-flipping boundary $\mathbf{B}_2$ . This can be achieved using Algorithm~\ref{alg:get_boundary_sign}. 
We assume that the middle of the sign-flipping boundary has an approximate value of $0$, because it has almost equal distance to two quantization boundaries.  

\circled{D} {Computing distance to sign-flipping boundary (second round EDT)} 
After sign-flipping boundaries $\mathbf{B}_2$ are obtained, we perform another round of EDT on $\mathbf{B}_2$ to compute all data points' distance to $\mathbf{B}_2$ (denoted as $\mathbf{Dist}_2$).
We omit computing the indices of the closest sign‑flipping boundary point to save memory and computation, as they will not be used in the interpolation (because the values of those points are assumed to be $0$). 

\circled{E} {Estimating and compensating errors} 
We then use two-point interpolation to estimate the error magnitude for each data point based on their distances to the quantization boundaries ($\mathbf{Dist}_1$) and sign-flipping boundary ($\mathbf{Dist}_2$). 
Assume that the $i$-th data point is non-boundary (both non-quantization-boundary and non-sign-flipping-boundary) and let $k_1=\mathbf{Dist}_1[i]$, $k_2=\mathbf{Dist}_2[i]$. We use the Inverse Distance Weighting (IDW) method to interpolate the $i$-th data point,
where the final compensation is computed as $C[i] = \frac{1/k_1}{1/k_1+1/k_2} S[i] \eta \epsilon$.
The $\eta$ here is a factor for error magnitude at the quantization boundary, and $\eta=1.0$ indicates that the error magnitude along the boundaries equals the error bound. 
In practice, the actual boundary error tends to be slightly lower than the error bound, so we set $\eta=0.9$ to account for possible variations in the boundary error magnitude. 
We have also validated this setting by an offline parameter sweep conducted on the evaluated datasets, where $\eta=0.9$ yields the best results most of the time. 
Due to space limitations, the detailed experimental results are not included.

\begin{algorithm}[t]
\caption{\textsc{GetBoundaryAndSignMap3D}} \label{alg:get_boundary_sign}
\footnotesize
\renewcommand{\algorithmiccomment}[1]{\hfill\textcolor{gray}{// #1}}
\begin{flushleft}
\textbf{Input}: \\
\quad Quantization index array $Q$ of size $N = d_0 \cdot d_1 \cdot d_2$ \\
\quad Dimensions $d_0$, $d_1$, $d_2$ \\
\textbf{Output}: \\
\quad $B$: 3D binary boundary map \\
\quad $S$: sign map at boundary locations
\end{flushleft}

\begin{algorithmic}[1]
\STATE Initialize $B[0 \dots N{-}1] \gets 0$ \COMMENT{Boundary map}
\STATE Initialize $S[0 \dots N{-}1] \gets 0$ \COMMENT{Sign map}
\FORALL{$i = 1$ to $d_0 - 2$ }
    \FORALL{$j = 1$ to $d_1 - 2$}
        \FORALL{$k = 1$ to $d_2 - 2$}
            \STATE $q \gets Q[i,j,k]$
            \IF{$q \ne$ any of its 6 neighbors}
                \STATE $B[i,j,k] \gets 1$ \COMMENT{Mark as quantization boundary}
                \STATE $S[i,j,k] \gets \texttt{sgn}(q_{\text{neighbor}} - q)$ \COMMENT{Get sign via forward difference}
                \STATE $\text{g}_x, \text{g}_y, \text{g}_z \gets$ \texttt{central\_diff}($i$, $j$, $k$, $Q$) \COMMENT{Compute gradient using central difference}
                \IF{$\max(\text{g}_x, \text{g}_y, \text{g}_z) \ge 1.0$}
                    \STATE $S[i,j,k] \gets 0$ \COMMENT{Fast-varying region—discard sign.}
                \ENDIF
            \ENDIF
        \ENDFOR
    \ENDFOR
\ENDFOR
\RETURN $ B, S $
\end{algorithmic}
\end{algorithm}

\begin{algorithm}[t]
\caption{\textsc{PropagateSignsAndConstructSignMap}} \label{alg:propsigns}
\footnotesize
\renewcommand{\algorithmiccomment}[1]{\hfill\textcolor{gray}{// #1}}
\begin{flushleft}
\textbf{Input}: \\
\quad $B$: boundary map where 1 indicates boundaries.\\ 
\quad $S_{B}$: sign map only initialized on boundaries.\\
\quad $I_{B}$: index of nearest boundary point for each location. \\
\textbf{Output}: \\
\quad $S$: Complete sign map.\\
\quad $B_2$: Sign-flipping boundary. 
\end{flushleft}
\begin{algorithmic}[1]
\STATE $S \gets S_{B}$      
\FOR{$i = 0$ to $N{-}1$}
    \IF{B[i] != 1 }
    \STATE $S[i] \gets S[I_{B}[i]]$ \COMMENT{In-place update}
    \ENDIF
\ENDFOR
\STATE $B_2 \gets \textsc{GetBoundary}(S)$
\RETURN $S , B_2$
\end{algorithmic}
\end{algorithm}

\begin{algorithm}[t]
\caption{\textsc{Distance-Based Compensation}} \label{alg:compensation}
\footnotesize
\renewcommand{\algorithmiccomment}[1]{\hfill\textcolor{gray}{// #1}}
\begin{flushleft}
\textbf{Input}: \\
\quad $D'$: decompressed data array \, 
\quad $Q$: quantization index array \\ 
\quad $N$: size of the data\,
\quad $\epsilon$: error bound\, 
\quad $\eta$: compensation factor \\
\textbf{Output}: \\
\quad $D''$: compensated decompressed data
\end{flushleft}
\begin{algorithmic}[1]
\STATE $B_1, S_{\text{B}} \gets \textsc{GetBoundaryAndSignMap3D}(Q)$
\STATE $Dist_{\text{1}}, I_{\text{1}} \gets \textsc{EuclideanDistanceTransform}(B_1)$
\STATE $S , B_2 \gets \textsc{PropagateSignsAndConstructSignMap}(B_1, S_{\text{B}},I_{\text{1}})$
\STATE $Dist_{\text{2}} \gets \textsc{EuclideanDistanceTransform}(B_2)$
\FOR{$i = 0$ to $N{-}1$}
    \STATE $dist_1 \gets Dist_{\text{1}}[i]$ \COMMENT{Distance  to boundary.}
    \STATE $dist_2 \gets Dist_{\text{2}}[i]$ \COMMENT{Distance to sign-flipping boundary.}
    \STATE $C \gets \textsc{Interpolate}(dist_1, dist_2, S[i], \eta\epsilon)$
    \STATE $D''[i] \gets D'[i] + C$ \COMMENT{Compensation added to decompressed  data.}
\ENDFOR
\RETURN $D''$
\end{algorithmic}
\end{algorithm}

\textbf{Full Algorithm and Complexity Analysis}: Putting everything together, we present our quantization-aware interpolation algorithm in Algorithm \ref{alg:compensation}. 
Note that we need to add the interpolated value to the quantized data to generate a representation close to the original data (line 8). 
The theoretical time complexity of this algorithm is $\mathcal{O}(N)$ because the complexity for all five steps is $\mathcal{O}(N)$. 

\section{Parallelization}\label{section:parallelization}

In this section, we present the parallelization of the proposed algorithm for both shared-memory and distributed-memory settings. We provide the CPU implementation as a proof-of-concept, along with a scaling study in this paper to demonstrate the scalability of our methods. The algorithm is inherently data-parallel and thus amenable to GPU acceleration, which we plan to explore in future work.

\subsection{Shared-memory Parallelization}
Our algorithm is highly parallelizable in shared-memory environments, as most steps can be parallelized in an embarrassingly parallel fashion.
For instance, Steps \circled{A} \circled{C} \circled{E} can be parallelized through collapsed loops. Although EDT computations in steps \circled{B} \circled{D} have strong dependencies along the processing dimension, they can be efficiently parallelized along the other dimensions. The parallel versions of the algorithm run in $\mathcal{O}(N/p)$  time with $p$ threads, demonstrating high efficiency and scalability. 

\subsection{Distributed-memory parallelization}
We also extend the parallelization of our algorithm to a distributed-memory setting using MPI.  This setting is more challenging due to strong dependencies inherent in the EDT processing. To address this challenge, we implement three strategies that consider the trade-off between performance and quality. 



\textit{Embarrassingly Parallel.} 
To achieve optimal scalability, our first implementation employs an embarrassingly parallel approach where each processor computes the entire algorithm locally without any inter-process communication.  
While this method maximizes scalability, it can lead to inaccuracies near process boundaries, as local information is not accounted for during computation. Thus, while theoretically optimal in scalability, this approach may result in reduced compensation quality.

\textit{Exact Parallelization (Sequentially-compliant)}.
Our second parallelization strategy rigorously implements the algorithm to make it compliant with the sequential implementation. However, this strategy incurs heavy communication overhead, leading to reduced scalability.  
Specifically, we need to exchange ghost elements to compute quantization boundaries, which involves stencil  communication with adjacent processors. 
Additionally,  when computing EDT along a column, processors on row $i$ cannot start unless all their preceding processors along the column complete their computation and send the data. 
This leads to very low parallelism with very high communication costs. 
Although this method provides the most accurate results, it suffers from the worst scalability.

\textit{Approximate Parallelization (Approximate algorithm with local communication)}.
To better balance the trade-off between performance and quality, we propose a lightweight communication strategy that only exchanges ghost elements for boundary computation, which only involves stencil communication. 
This strategy emphasizes that boundary information most significantly impacts adjacent elements, requiring minimal communication. 
With this approach, only two rounds of stencil communication are needed for Steps \circled{A} \circled{C}, which allow for improved scalability.

\begin{figure}
    \centering
\includegraphics[width=0.95\linewidth]{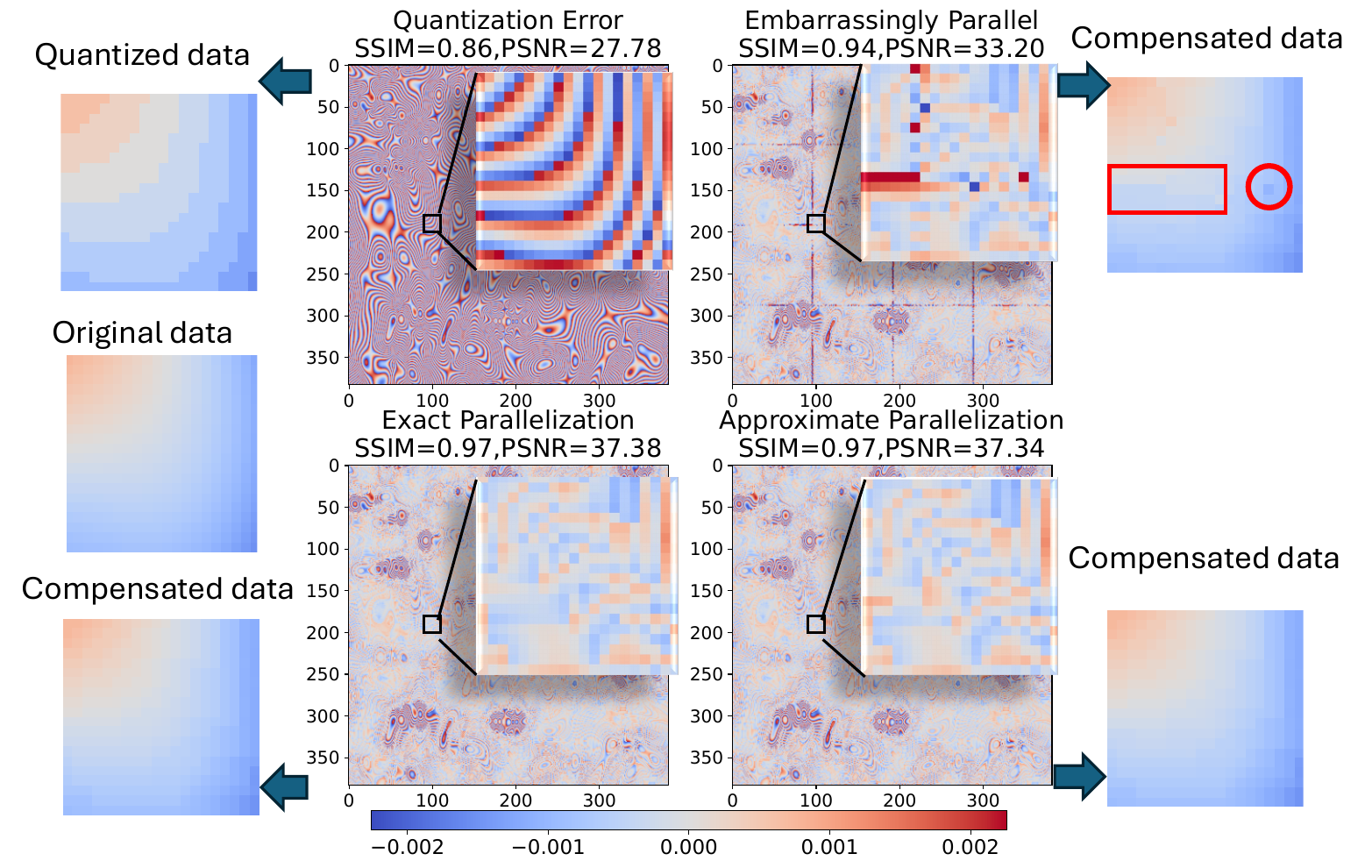}
    \caption{Visualization of error slices for the three parallel methods. Original data, decompressed data and compensated decompressed data are shown on two sides respectively (visualization range is [-0.032, 0.032]).}
    \label{fig:error-vis-blocks}
    \vspace{-\baselineskip}
\end{figure}

We showcase the quality of the three parallelization strategies using a toy example with 64 cores on 3D data. 
Fig.~\ref{fig:error-vis-blocks} visualizes the error and data distribution for the quantized data (top left) and compensated data (bottom left) using the three communication patterns. 
Local window SSIM and PSNR are reported here. 
Notably, the embarrassingly parallel implementation exhibits striping patterns in the compensated error of the compensated data, leading to visible artifacts. In contrast, the compensated data produced by the other two parallelization methods demonstrate high quality, as well as improved SSIM and PSNR metrics.

\section{Evaluation}\label{section:evaluation}


To verify the effectiveness and performance of our methods, we design and conduct systematic experiments. 

\subsection{Experiment setup}
We conduct the experiments on a medium-sized cluster~\cite{anonymous_cluster}. Each node has two 64-core AMD EPYC 7702 CPUs and 512 GB of DDR4 memory. The software environment includes CentOS~8.4, GCC~9.3, and Intel MPI~2021.2.0.  In order to evaluate our methods comprehensively and systematically, five datasets from different application domains are used for the experiments. They are listed in Table~\ref{tab:dataset-information}. We include CESM from Earth Climate Research~\cite{kay2015community}, 
Hurricane from weather simulation~\cite{hurricane-data}, NYX from cosmological simulation~\cite{nyx}, S3D from combustion simulation~\cite{chen2017s3d}, and JHTDB from turbulence simulation \cite{data-jhtdb}.
The first four datasets are used for small-scale tests, including rate-distortion experiments and shared-memory parallelization performance evaluation; the results are reported as the aggregation across the different fields. The JHTDB dataset is used to demonstrate the MPI implementation’s performance in terms of quality metrics and throughput. 

\textit{Baseline:} We use three classical and efficient filters widely applied in image restoration: the Gaussian filter, the uniform filter, and the Wiener filter~\cite{gaussian_image_restore_ramadan2018optimum,wiener_image_restore, gaussian_wiener_ramadan2019effect}. These filters can be readily adapted to scientific data while maintaining reasonable processing speed. We exclude neural network–based artifact mitigation methods from this study for two main reasons: (1) suitable pretrained models for our data modality are not readily available, and (2) training custom networks would require substantial data curation and computational resources. Furthermore, neural network inference is typically more resource-intensive and less portable for our targeted workflows. 




\begin{table}[h]
    \centering
    \footnotesize
    \caption{Dataset Information}
    
    \small
\resizebox{0.99\columnwidth}{!}{  
    \begin{tabular}{|c|c|c|c|c|}
    \hline
    Application &\# Fields& Dimension & Total Size& Domain\\
    \hline
    CESM\cite{kay2015community} & 77 & $1800\times3600$ & 1.86 GB& Climate \\
    \hline 
    Hurricane\cite{hurricane-data}& 13 & 100$\times$500$\times$500 &1.2 GB& Weather\\
    \hline
    NYX\cite{nyx} & 6 & 512$\times$512$\times$512 &3.1 GB& Cosmology\\
    \hline 
    S3D \cite{chen2017s3d}& 11 & 500$\times$500$\times$500 &5.12 GB& Combustion\\
    \hline
    JHTDB\cite{data-jhtdb} & 1 & 4096$\times$4096$\times$4096 & 256 GB&Turbulence\\
    \hline
    \end{tabular}
    }
    \label{tab:dataset-information}
\end{table}

\subsection{Evaluation configurations and metrics}

In the evaluations, we use a value-range-based relative error bound (EB), obtained by dividing the absolute error bound by the field’s value range.  We overload $\epsilon$ to denote this relative bound in the remaining sections. This convention is widely used in the compression community \cite{sz16,sz17,sz18}. The following metrics are used.  

\begin{itemize}[leftmargin=*]
\item Maximum error: the maximum absolute (or relative) error between the original data and the compensated data, used to evaluate error control.

\item Bit-rate: the average number of bits required to represent a single data point in the compressed file, calculated as bitwidth/compression ratio (bitwidth is 32 for single-precision data). 

\item Throughput and parallelization efficiency: used to evaluate performance and scalability. 
\item EB-Distortion: plots showing EB used for quantization versus the distortion metrics (i.e., PSNR or SSIM in this evaluation) of the quantized and compensated data. 
\item Rate-Distortion: plots showing compressed bit-rate versus PSNR or SSIM of the quantized and compensated data. 
\end{itemize}

\subsection{Guaranteed Error Control with Relaxed Error Bound}
Table~\ref{tab:error-control-gaussian} reports the maximum relative errors observed across multiple datasets and fields after applying Gaussian filtering ($\sigma =1.0$), Uniform filtering, Wiener filtering, and our compensation method. We use a $3\times3\times3$ window for all three filters. For fair comparison, we use the estimated variance $\epsilon^2/3$ for Wiener filtering noise power because the true variance is unknown at post-decompression time. We compare the maximum relative errors with a relaxed error bound $(1+\eta)\epsilon$ (see Section~\ref{section:method} and Algorithm \ref{alg:compensation}). 
 
The smoothing filters (Gaussian, Uniform, Wiener) reduce posterization artifacts but can introduce larger errors near boundaries or sharp transitions. The results show that the maximum error after filtering often exceeds the relaxed error bound, meaning these filters do not strictly guarantee error control. 
Wiener filtering typically yields the best error control among the filters given the estimated variance, but compliance with the relaxed bound varies across datasets and fields.  In contrast, by using $\eta=0.9$ as the compensation  factor, our method ensures that the resulting errors remain strictly bounded by the relaxed error bound.



\begin{table}[h]
\caption{Maximum Relative Error After Compensation ($\epsilon = 1\times10^{-3}$)}\label{tab:error-control-gaussian}
\centering
\footnotesize
\begin{tabular}{|c|c|c|c|c|c|}
\hline
Dataset & Field & Gaussian & Uniform & Wiener & Ours \\ \hline
\multirow{2}{*}{CESM} & CLDHGH & 0.0868 & 0.081 & 0.0024 & 0.0019 \\ \cline{2-6} 
 & CLDLOW & 0.129 & 0.1218 & 0.0023 & 0.0019 \\ \hline
\multirow{2}{*}{Hurricane} & Uf48 & 0.2809 & 0.2814 & 0.0016 & 0.0019 \\ \cline{2-6} 
 & Wf48 & 0.5456 & 0.5592 & 0.0022 & 0.0019 \\ \hline
\multirow{2}{*}{NYX} & temperature & 0.3804 & 0.3614 & 0.0017 & 0.0019 \\ \cline{2-6} 
 & velocity\_x & 0.191 & 0.1746 & 0.002 & 0.0019 \\ \hline
\multirow{2}{*}{S3D} & field0 & 0.0188 & 0.0147 & 0.0011 & 0.0019 \\ \cline{2-6} 
 & field10 & 0.1756 & 0.1756 & 0.0013 & 0.0019 \\ \hline
\end{tabular}
\end{table}


\subsection{Rate-distortion}
\begin{figure}[h]
    \centering
    \includegraphics[width=\linewidth]{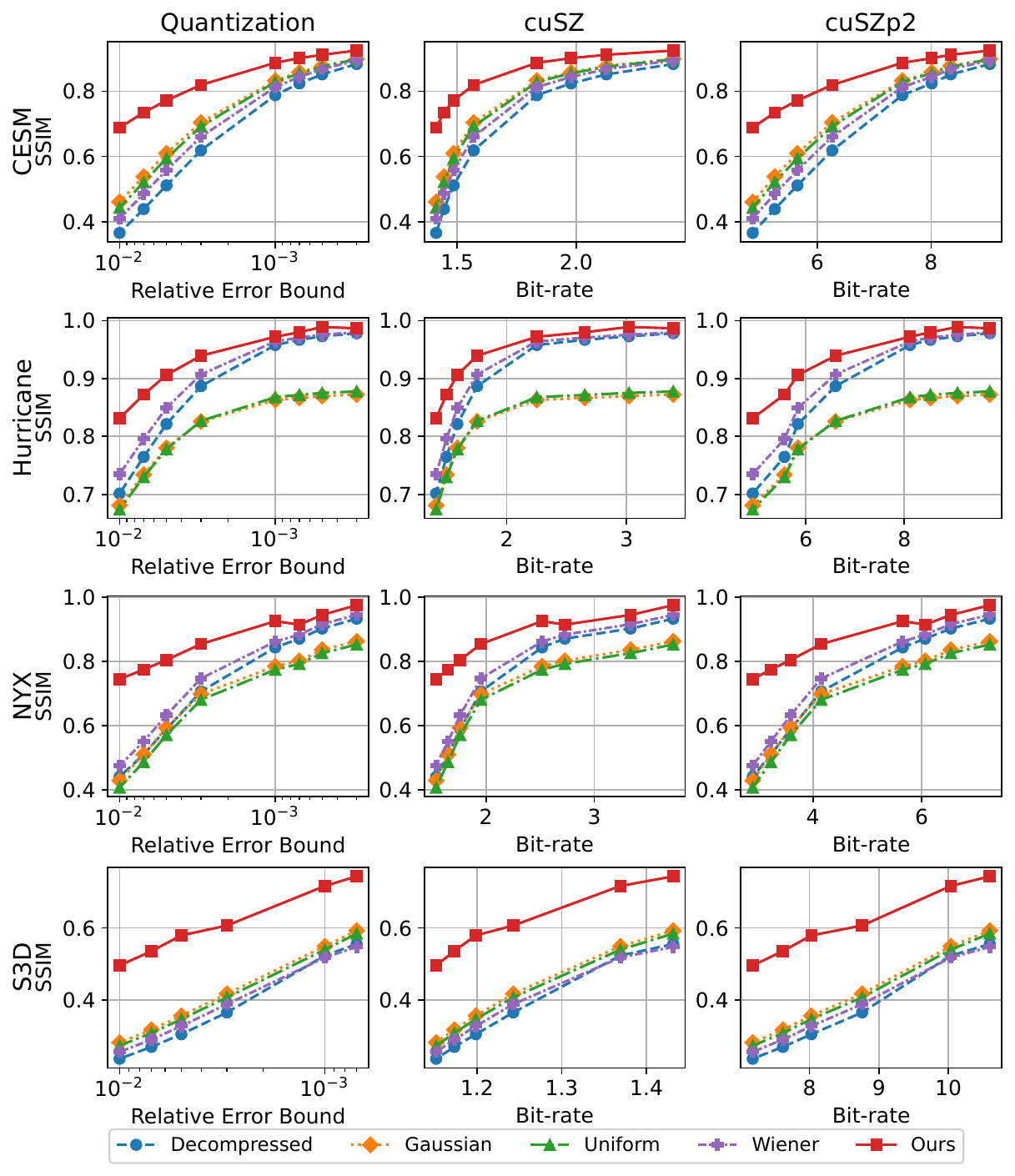}
    \caption {Rate-distortion results with SSIM. }
    \label{fig:ssim-br}
\end{figure}

\begin{figure}[h]
    \centering
    \includegraphics[width=\linewidth]{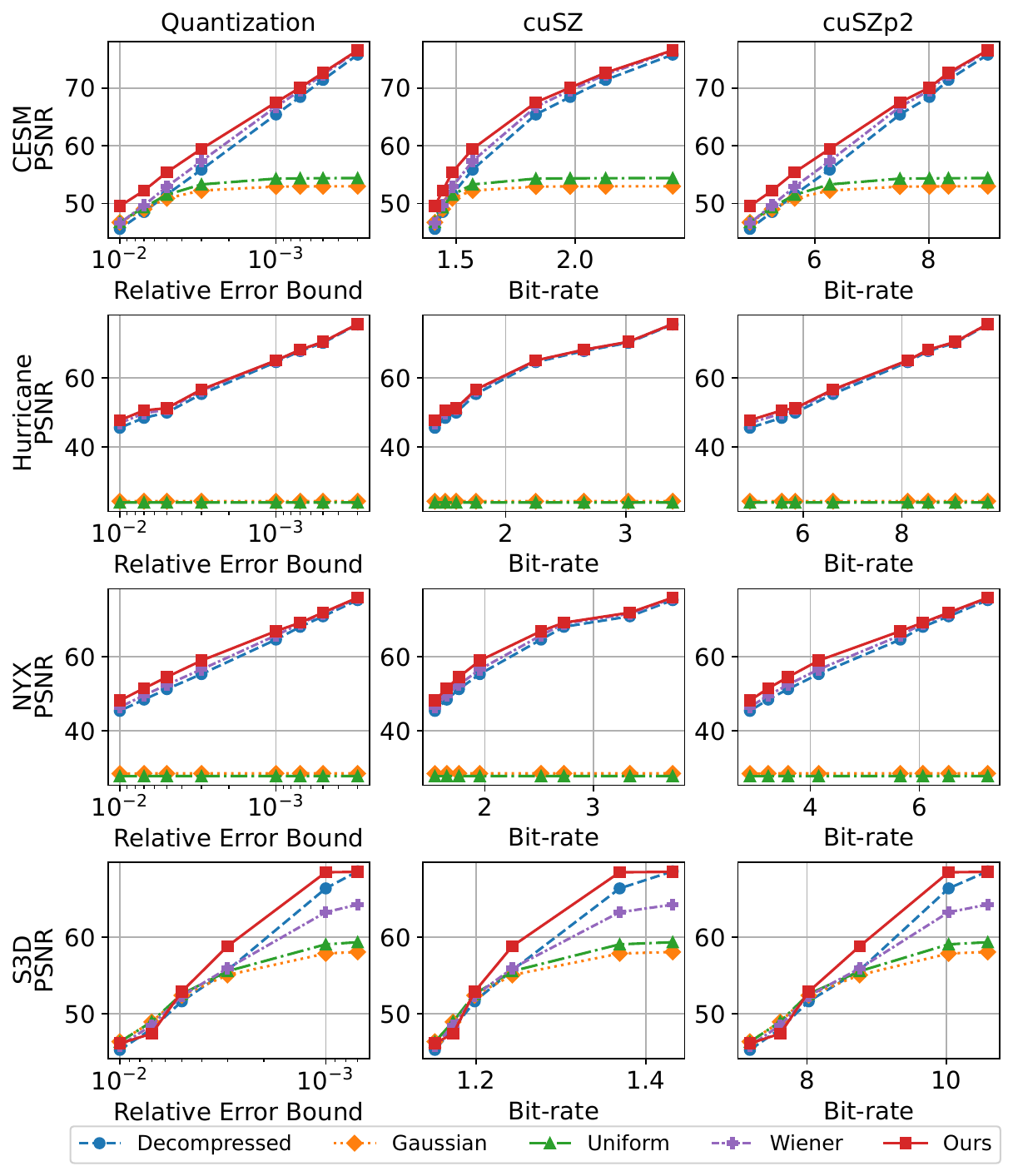}
\vspace{-1em}
    \caption {Rate-distortion results with PSNR. }
    \label{fig:psnr-br}
\end{figure}

The rate-distortion results are presented in Fig.~\ref{fig:ssim-br} and Fig.~\ref{fig:psnr-br}. The first column of each figure shows SSIM and PSNR as functions of the prescribed relative error bound. We compare the following approaches: (1) quantized/decompressed data, (2) Gaussian-filtered quantized data, (3) Uniform-filtered quantized data, (4) Wiener-filtered quantized data, and (5) our compensation method. 

Overall, our compensation method consistently improves SSIM across all four datasets without degrading PSNR. Gaussian and Uniform filtering sometimes improve SSIM but can significantly degrade PSNR. Wiener filtering generally yields SSIM improvements with less PSNR degradation than Gaussian or Uniform filters.

On CESM, the SSIM increase rate ranges from 4.68\% to 88.20\%. The largest SSIM increase (from 0.36  to 0.79) appears when $\epsilon = 0.01$, where bit-rate for cuSZ and cuSZp2 are 1.41 and 4.86, respectively. At this point, PSNR increases by 8.64\%. If we target the same SSIM level, the compression ratio gains are 1.19$\times$ and 1.39$\times$ for cuSZ and cuSZp2, respectively.  On Hurricane, the SSIM increase rate ranges from 0.75\% to 16.94\%, with the largest  SSIM increase at  $\epsilon =0.01$. On NYX, the SSIM increase rate ranges from 4.27\% to 48.93\%, and the largest SSIM increase appears at $\epsilon=0.01$.

On the S3D dataset, we observe the largest SSIM improvement across all datasets: an increase of 108.33\%, from 0.24 to 0.49, at $\epsilon = 0.01$. This corresponds to up to 1.17$\times$ and 1.34$\times$ gain in compression ratios for cuSZ and cuSZp2, respectively, when SSIM is matched. We also observe a slight PSNR decline when $\epsilon = 0.001$ and the underlying cause is discussed in the following case study.

\begin{figure}[h]
    \centering
    \includegraphics[width=0.95\linewidth]{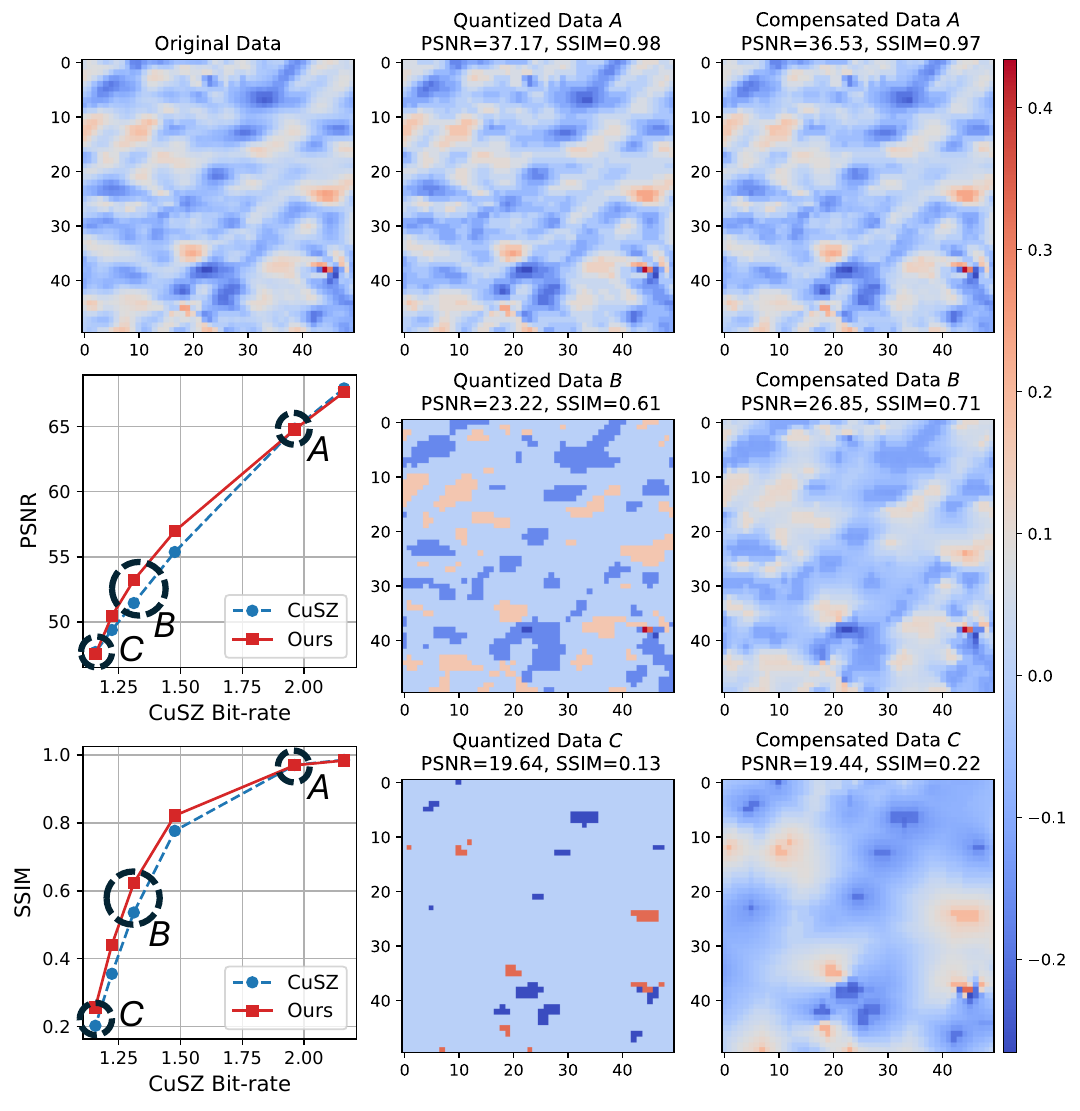}
\vspace{-1em}
    \caption {Hurricane-Wf48 study (local PSNR/SSIM reported on visualization).}
    \label{fig:hurricane-case-study}
\end{figure}

\textbf{Visualization case study:} 
We present a visual case study to analyze the effectiveness of our method under different compression error bounds. 
We use the Wf48 field from the Hurricane dataset as an example, and the results are displayed in Fig.~\ref{fig:hurricane-case-study}. 
Specifically, we select three points from the rate-distortion curves for visualization, which correspond to different error bounds. At point \textbf{\textit{A}}, where the error bound is relatively low, the quantized data has almost no visual artifacts. Our method yields little improvement in this case, but it does not incur degradation either. At point \textbf{\textit{B}}, where the error bound is moderate, our method can significantly improve the visual quality, PSNR, and SSIM. At point \textbf{\textit{C}}, where the error bound is very high, our improvement over the quantized data is minor in terms of PSNR due to the limited availability of meaningful information. Meanwhile, our improvement in SSIM is still noticeable, as our interpolation mechanism is expected to have better perceptual quality than the quantized data. 
These results are consistent with the rate-distortion curves in Figs.~\ref{fig:ssim-br} and~\ref{fig:psnr-br}, and they indicate that our method works best at moderate error bounds, where artifacts are present with a rough silhouette of the data. 
Note that moderate error bounds are the most common use cases for scientific data compressors, because a low/high error bound typically yields limited compression ratios or unusable data quality, respectively.



\subsection{Performance Evaluation}

\subsubsection{OpenMP Scalability}

Since our method operates on decompressed data, we compare the efficiency of our CPU-based shared-memory implementation with the decompression performance of two widely used CPU-based compressors that support OpenMP: SZp and SZ3. SZp emphasizes high throughput and has been employed in homomorphic compression and communication frameworks~\cite{huang2024hzccl}, while SZ3~\cite{zhao2021optimizing} balances compression ratio and throughput, and has been applied in domains such as reverse time migration~\cite{yanfan_rtm_sz3} and Morse–Smale segmentation correction~\cite{msz_yuxiao}. Importantly, our method can be seamlessly integrated into the decompression pipelines of both SZp and SZ3, making this efficiency comparison directly relevant to potential integration implementations.

\begin{figure}[h]
    \centering
    \includegraphics[width=0.9\linewidth]{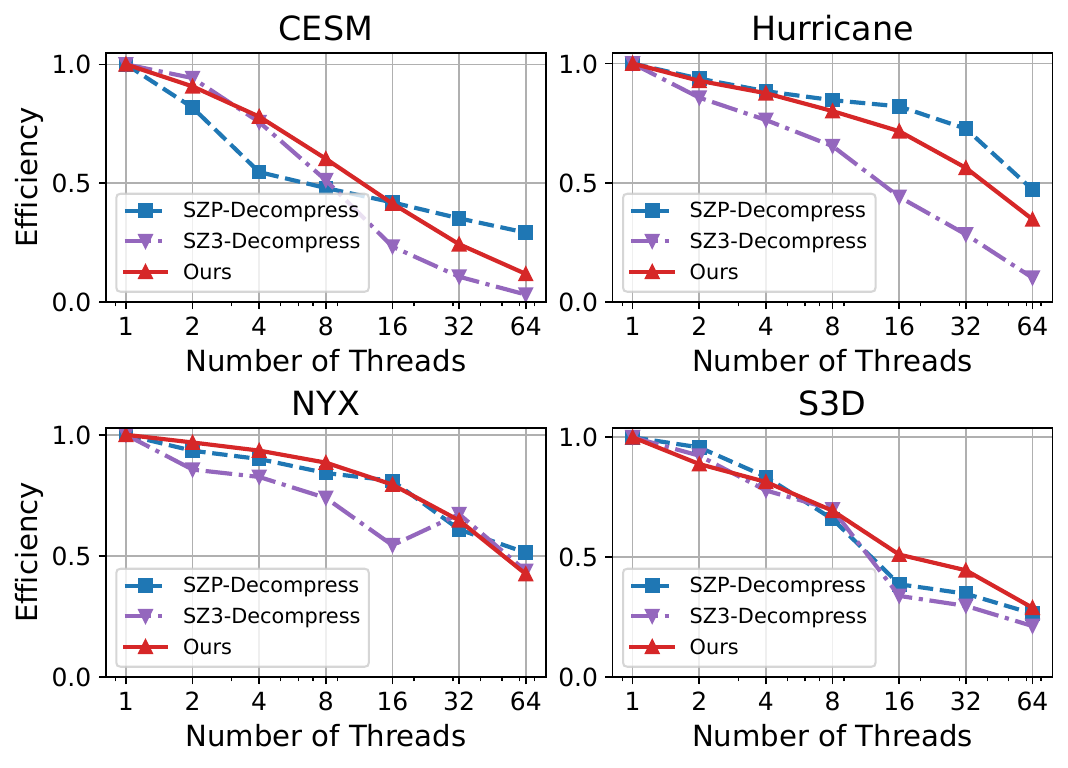}
\vspace{-1em}
    \caption {Shared memory efficiency($\epsilon=0.001$).}
    \label{fig:omp_efficiency}
\end{figure}

The results are shown in Fig.~\ref{fig:omp_efficiency}. 
Efficiency is calculated as the multi-thread speedup relative to the number of threads. Our findings indicate that the efficiency of our method is competitive with that of SZp, suggesting that we are able to maintain high scalability in a shared-memory setting. Furthermore, we observe that our approach outperforms SZ3 in most scenarios. Notably, we see a decreasing trend in the efficiency curves, which is attributed to the overhead associated with increasing the number of threads. Additionally, efficiency is influenced by the size and characteristics of the data.

\subsubsection{Throughput and quality of MPI parallelization}

In this section, we use JHTDB for experiments, which contains the velocity field for a turbulence simulation. The total size is 256 GB and the original data dimension is $4096\times4096\times 4096$. To comprehensively assess the scalability of the proposed methods, we design two setups for distributed-memory parallelization experiments.

\begin{figure}[h]
    \centering
\includegraphics[width=1\linewidth]{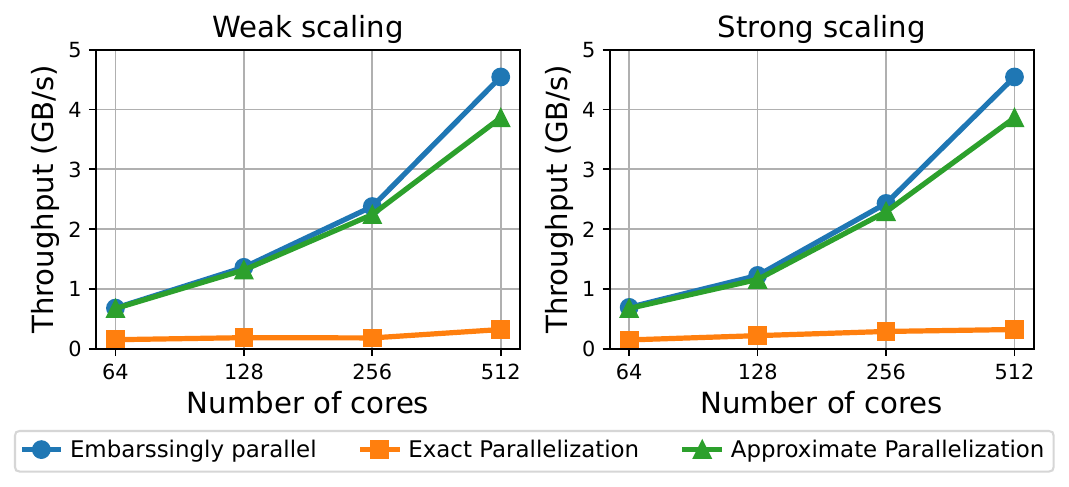}
\vspace{-1em}
    \caption{Distributed memory throughput.}
    \label{fig:mpi-performance}
\end{figure}

\textbf{Weak scaling}: In this setup, each core processes the same data size of $512\times512\times512$. To balance the workload, we distribute 64, 128, 256, and 512 tasks evenly across 2, 4, 8, and 16 nodes, respectively. The results are shown in Fig.~\ref{fig:mpi-performance}. For the 128-core, 256-core, and 512-core settings, the efficiency of the Approximate Parallelization is $100\%$, $87.5\%$ and $83.6\%$ with respect to the 64-core base case, while the efficiency for Embarrassingly Parallel is $97.9\%$, $83.4\%$, and $71.8\%$.  The Exact Parallelization has the lowest throughput  and the worst scalability across the three settings. At the 64-core setting, it shows 154.36 MB/s throughput, while Embarrassingly Parallel and Approximate Parallelization show 695.93 MB/s and 688.48 MB/s, respectively. 


\textbf{Strong scaling}: The original data is decomposed into the following dimensions: 
$1024\times1024\times 1024$,
$1024\times1024\times 512$, 
$1024\times512\times 512$, 
$512\times 512 \times 512$. For each of these decompositions, we distribute the tasks evenly across 16 nodes, using 64, 128, 256, and 512 tasks, respectively. Fig.~\ref{fig:mpi-performance} shows the throughput results of these experiments. In this setup, the efficiency of the 128-core, 256-core, and 512-core settings for  Approximate Parallelization is 86.0\%, 85.0\% and 71.1\% with respect to the 64-core base case. In contrast, the Embarrassingly Parallel approach shows 88.6\%, 87.9\% and 82.2\% efficiency under the same settings.  The Exact Parallelization has poor throughput and scalability compared to the other two communication patterns due to the use of sequential computation for a global EDT. This highlights the scalability limitation of a globally sequential EDT in a distributed-memory setting.

\begin{figure}[h]
\centering
\vspace{-1em}
\includegraphics[width=1\linewidth]{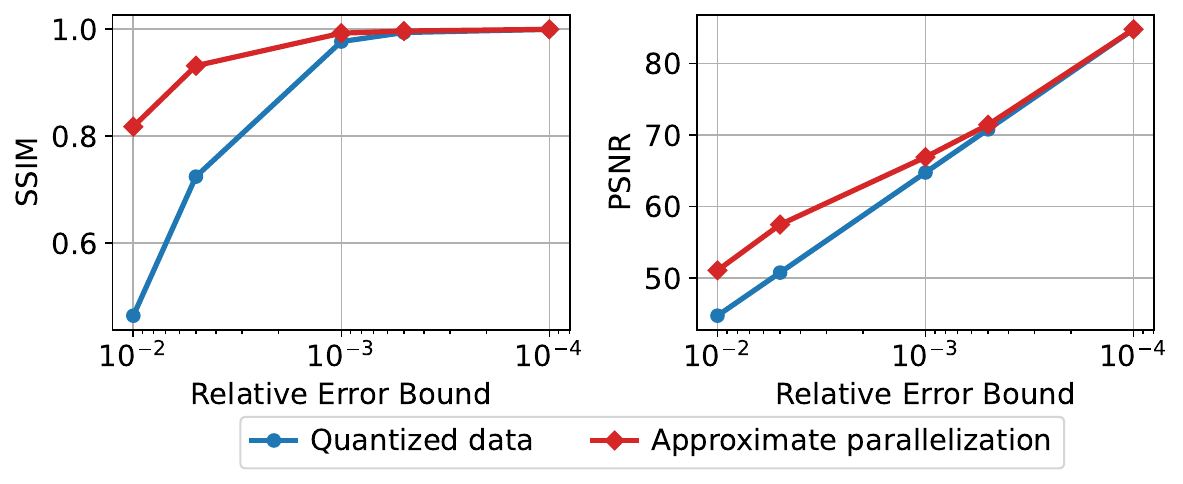}
\vspace{-1em}
\caption{JHTDB EB-distortion results. }
\label{fig:jhtdb-rate-distortion}
\end{figure}

\textbf{Quality evaluation}: Fig.~\ref{fig:jhtdb-rate-distortion} presents the EB-Distortion results for the JHTDB dataset. The data is decomposed into 512 of $512\times512\times512$ blocks and processed using Approximate Parallelization across 512 cores. We observe a maximum SSIM increase rate of $76.02\%$ and up to a $14.12\%$ increase in PSNR when $\epsilon=0.01$. 

\textbf{Overhead discussion:} The efficiency loss in the communication-free Embarrassingly Parallel approach is attributed to load imbalance: per-rank execution times ranged from 38s to 50s for the 256-rank configuration, with the wall-clock time determined by the slowest rank after barrier synchronization. This variation reflects the heterogeneous computational complexity across different regions of the turbulence data. This imbalance is also attributed to the efficiency loss in the Approximate Parallelization approach. 

\begin{figure}[h]
    \centering
\includegraphics[width=0.85\linewidth]{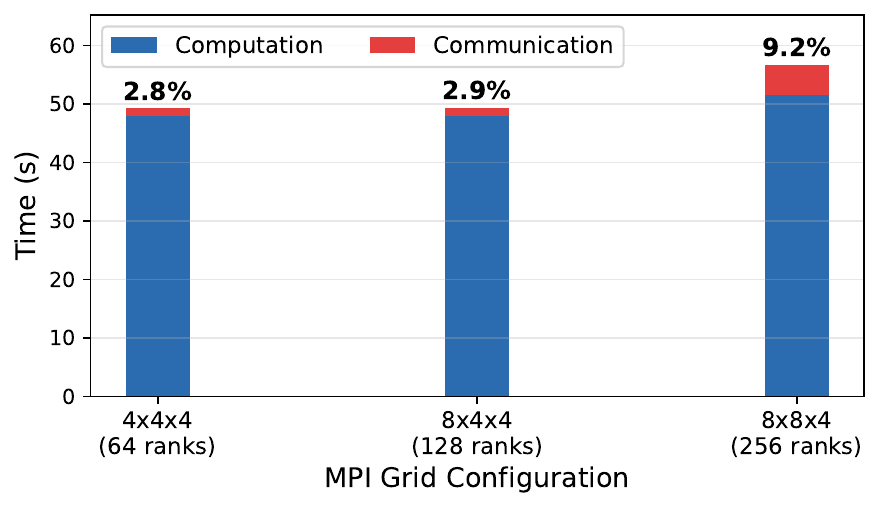}
\vspace{-1em}
    \caption{Breakdown of execution time into computation and communication.}
    \label{fig:comm-breakdown}
\end{figure}

We further analyze the communication cost of the Approximate Parallelization strategy under 64, 128, and 256 ranks. Fig.~\ref{fig:comm-breakdown} presents the time breakdown for weak scaling, where the proportion of communication time is denoted. We run the program for 5 iterations and report the average total time after discarding the maximum and minimum values. The communication time is reported as the sum of the two data exchange rounds in steps \circled{A} and \circled{C}. 
With 64 and 128 ranks, the average exchange time remains stable at approximately 1.4\,s ($<$3\% of total time), demonstrating that the algorithm scales well across 2 - 4 nodes. However, the exchange time rises to 5.24\,s (9.25\%) with 256 ranks (8 nodes). This increase is caused by~(1) the synchronization overhead from load imbalance mentioned above, where faster-finishing ranks idle-wait at the halo exchange boundary until their slower neighbors complete computation, and ~(2) system-level resource contention that leads to a large variation in communication time, as we observe that the communication time fluctuates between 2.56s and 5.84s across the 5 test iterations.


\section{Conclusion}\label{section:conclusion}
In this paper, we present an efficient algorithm to mitigate artifacts in pre-quantization based compressors. In particular, we first characterize such artifacts to understand correlations in compression errors, and then propose a quantization-aware interpolation algorithm to approximate the compression errors, which are then added to the decompressed data for artifact mitigation. We further parallelize our algorithm using both OpenMP and MPI to achieve high throughput and parallel efficiency. Across five diverse scientific datasets, the proposed method consistently enhances decompressed data quality, achieving up to 108.33\% improvement in SSIM over the unmitigated output of pre-quantization based compressors.

Future work will develop adaptive strategies for regions with homogeneous quantization indices or extremely large error bounds. We will implement and optimize the algorithm on GPUs to achieve higher throughput.  



\section*{Acknowledgment}

The material was supported by the U.S. Department of Energy, Office of Science, Advanced Scientific Computing Research (ASCR), under contract DE-AC02-06CH11357. This material was also supported by the National Science Foundation under Grant Nos. OAC-2311875, OAC-2104023, OAC-2313122, OAC-2504255, and OAC-2504254.
We would like to thank the University of Kentucky Center for Computational Sciences and Information Technology Services Research Computing for its support and use of the Lipscomb Compute Cluster, Morgan Compute Cluster, and associated research computing resources.

\bibliographystyle{IEEEtran}
\bibliography{sample-base}

\end{document}